\documentclass[twocolumn,superscriptaddress,floatfix,
longbibliography,aps,pra,preprintnumbers]{revtex4-1}

\usepackage[utf8]{inputenc}
\usepackage[T1]{polski}
\usepackage[english]{babel}

\usepackage{graphicx}
\usepackage{bm}
\usepackage{color}
\usepackage{epstopdf}
\usepackage{amsmath}
\usepackage{amssymb}
\usepackage{epstopdf}
\usepackage[normalem]{ulem}
\usepackage{nicefrac,xfrac}

\usepackage[urlcolor=blue,colorlinks=true,citecolor=blue,linkcolor=blue,pdfstartview={FitH},bookmarks=false]{hyperref}

\graphicspath{{fig/}{./fig/}{.}}

\begin{document}

\title{First principles study of topological phase in chains of $3d$ transition metals}

\author{Aksel Kobia\l{}ka}
\email[e-mail: ]{akob@kft.umcs.lublin.pl}
\affiliation{\mbox{Institute of Physics, Maria Curie-Sk\l{}odowska University,
Plac Marii Sk\l{}odowskiej-Curie 1, PL-20031 Lublin, Poland}}

\author{Przemys\l{}aw Piekarz}
\email[e-mail: ]{piekarz@wolf.ifj.edu.pl}
\affiliation{\mbox{Institute of Nuclear Physics, Polish Academy of Sciences,
ulica W. E. Radzikowskiego 152, PL-31342 Krak\'{o}w, Poland}}

\author{Andrzej M. Ole\'{s}}
\email[e-mail: ]{a.m.oles@fkf.mpg.de}
\affiliation{\mbox{Institute of Theoretical Physics, Jagiellonian University,
Profesora Stanis\l{}awa \L{}ojasiewicza 11, PL-30348  Krak\'{o}w, Poland}}
\affiliation{Max Planck Institute for Solid State Research,
Heisenbergstrasse 1, D-70569 Stuttgart, Germany}

\author{Andrzej Ptok$\,$}
\email[e-mail: ]{aptok@mmj.pl}
\affiliation{\mbox{Institute of Nuclear Physics, Polish Academy of Sciences,
ulica W. E. Radzikowskiego 152, PL-31342 Krak\'{o}w, Poland}}

\date{\today}

\begin{abstract}
Recent experiments have shown the signatures of Majorana bound states at the ends of magnetic chains deposited on a superconducting substrate.
Here, we employ first principles calculations to directly investigate the topological properties of $3d$ transition metal nanochains ({\it i.e.}, Mn, Cr, Fe and Co).
In contrast to the previous studies [Nadj-Perge {\it et al.}
\href{http://doi.org/10.1126/science.1259327}{Science {\bf 346}, 602 (2014)}
and Ruby {\it et al.}
\href{http://doi.org/10.1021/acs.nanolett.7b01728}{Nano Lett. {\bf 17}, 4473 (2017)}],
we found the exact tight binding models in the Wannier orbital basis 
for the isolated chains as well as for the surface--deposited wires.
Based on these models, we calculate topological invariant of $\mathbb{Z}_2$ 
phase for all systems.
Our results for the isolated chains demonstrate the existence of the 
topological phase only in the Mn and Co systems. We considered also a 
non-collinear magnetic order as a source of the non--trivial topological 
phase and found that this type of magnetic order is not a stable ground 
state in the Fe and Co isolated chains. Further studies showed that a 
coupling between the chain and substrate leads to strong modification of 
the band structure. Moreover, the analysis of the topological invariant 
indicates a possibility of emergence of the topological phase in all 
studied nanochains deposited on the Pb surface. Therefore, our results 
demonstrate an important role of the coupling between deposited atoms 
and a substrate for topological properties of nanosystems.
\end{abstract}


\maketitle

\section{Introduction}
\label{sec.intro}

Prediction of localization of the Majorana bound states (MBS) at  the ends of the one-dimensional chain~\cite{kitaev.01} initiated intensive studies of this phenomenon in wide array of systems~\cite{beenakker.13,stanescu.tewari.13,beenakker.15,lutchyn.bakkers.18}.
Typically, to generate MBS a mutual interplay between the conventional {\it s-wave} superconductivity, Zeeman magnetic field and strong spin-orbit coupling is essential~\cite{oreg.refael.10,lutchyn.sau.10}.
This condition can be achieved in semiconductor--superconductor
nanostructures, where a semiconducting nanowire is deposited on a
conventional superconductor \cite{mourik.zuo.12,deng.yu.12,das.ronen.12,finck.vanharlingen.13,lee.jiang.13,deng.vaitiekenas.16,albrecht.higginbotham.16,nichele.drachmann.17}.
Other theoretically predicted possibilities of the emergence of MBS, are chains of the magnetic atoms~\cite{nadjperge.drozdov.13,li.chen.14,peng.pientka.15,pawlak.silas.19}
or nanoparticles~\cite{choy.edge.11} located on a superconductor.
The interplay between the magnetic moments and proximity induced
superconductivity can drive the system into a topological phase
\cite{braunecker.simon.13,klinovaja.stano.13}.

Scanning tunneling microscopy (STM) technique has been proven to be
an excellent tool in this venue. The experiment based on the
theoretical prediction was carried out in 2014 by Yazdani group
\cite{nadjperge.drozdov.14} -- the authors presented the evidence of 
forming of topological Majorana zero modes in iron chains on the
superconducting Pb(110) surface.
Additionally, high-resolution experiments with superconducting tips confirmed the existence of zero-energy excitations in this type of chain~\cite{ruby.pientka.15} and also, in the form of zero-energy
local density of states (LDOS) measurement~\cite{pawlak.kisiel.16}.
More recently, the spin-dependent experiments~\cite{wiesendanger.09} demonstrated the emergence of the MBS in this system~\cite{jeon.xie.17}.

\begin{figure}[!b]
\centering
\includegraphics[width=\linewidth]{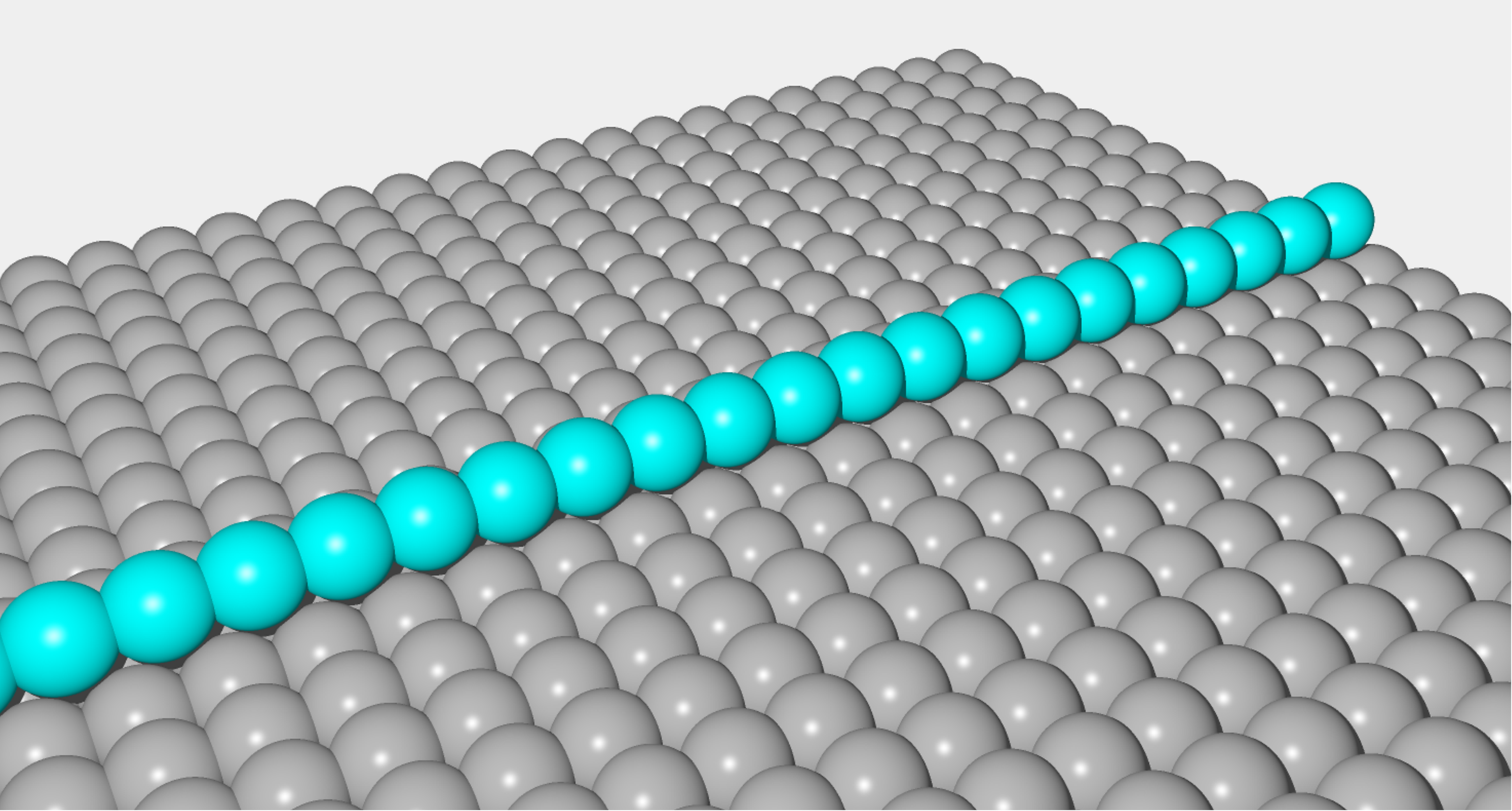}
\caption{
Schematic representation of a discussed system: the monoatomic magnetic
chain of $3d$ transition metal (cyan atoms) at the surface of
superconductor (gray atoms).
\label{fig.schem}
}
\end{figure}

The mentioned experiments are based on the existence of the Yu--Shiba--Rusinov (YSR) in-gap bound states induced by a magnetic impurity~\cite{yu.65,shiba.68,rusinov.69}.
The interaction of the local spin of impurity with the Cooper pairs in superconductor gives rise to a low-lying excited state within the gap of the quasiparticle excitation spectrum~\cite{balatsky.vekhter.06,heinrich.pascual.18}.
Progress in experimental techniques allows the study of the YSR bound states of individual magnetic atoms~\cite{yazdani.jones.97}. Such studies of the YSR bound states were performed for many $3d$ transition metal adatoms, like {\it e.g.} Ti~\cite{yang.bae.17},
Mn~\cite{yazdani.jones.97,moca.demler.08,ruby.peng.16,ruby.heinrich.18,ji.zhang.08,ji.zhang.10}, Cr~\cite{ji.zhang.08,ji.zhang.10,choi.rubioverdu.17,choi.fernandez.18},
Fe~\cite{menard.guissart.15,cornils.kamlapure.17}, or
Co~\cite{meier.zhou.08}. Forming the chain of magnetic adatoms can lead to the evolution of the YSR bound states to the zero--energy MBS~\cite{chevallier.simon.13,pientka.glazman.13,andolina.simon.17}.
In case of the chains of magnetic atoms, the rich spectrum of the in-gap states can be observed~\cite{bjornson.balatsky.17,mohanta.kampf.18}.

Experimentally, the monoatomic chains are usually prepared by the electron beam evaporation technique.
This method was used successfully in the case of the Fe~\cite{nadjperge.drozdov.14,ruby.pientka.15,feldman.randeria.16,pawlak.kisiel.16,jeon.xie.17} and Co~\cite{ruby.heinrich.17} chains.
However, recent progress in atomic engineering~\cite{eigler.schweizer.90,khajetoorians.wiebe.12,morgenstern.lorente.13,spinelli.rebergen.15,choi.lorente.19,schneider.brinker.20} allows for \textit{in situ} construction of the magnetic atomic chains~\cite{hirjibehedin.lutz.06,matsui.meyer.07,folsch.yang.09,loth.baumann.12,yan.malavolti.17,rolfpissarczyk.yan.17,kim.palaciomorales.18,steinbrecher.rausch.18,kim.palaciomorales.18,kamlapure.corils.18}.
This technique can help to produce monoatomic chains on the superconducting surfaces (Fig.~\ref{fig.schem}).
In relation to chains prepared by electron beam technique~\cite{nadjperge.drozdov.14,ruby.pientka.15,feldman.randeria.16,pawlak.kisiel.16,jeon.xie.17,ruby.heinrich.17}, artificial magnetic chains can have predetermined parameters, such as distance between atoms.
Additional advantage of this technique is a possibility for preparation of ideally homogeneous system.
By pushing this idea further, the pristine, homogeneous chains of 40 atoms and longer were produced~\cite{kamlapure.corils.18} by
Wiesendanger group, using an \textit{in--situ} STM assembly
\cite{kim.palaciomorales.18}. The zero-energy MBS at the Fe chain ends became more stable with increase of the nanochain length.

In the context of the mentioned experiments, in this paper we study the physical properties of monoatomic chains of magnetic $3d$ transition metal atoms, {\it i.e.}, Mn, Cr, Fe and Co.
Our studies take advantage of first principles calculations and the parameters obtained using this method are applied in a tight binding model (TBM) in order to calculate topological invariants of the investigated systems.
In the previous studies, the analysis of topological properties of monoatomic chains was based on tight-binding models with the hopping parameters taken from bulk crystals~\cite{nadjperge.drozdov.14,ruby.heinrich.17}.
Since the electronic band structures of monoatomic chains significantly differ from those of crystals, such simplified approach may lead to wrong conclusions.
Surprisingly, our calculations for the $3d$ monoatomic chains show that a non--trivial topological phase may exist only in Mn and Co free-standing nanowires, while this phase is excluded in Fe or Cr chains.
These results for isolated chains are incompatible with previous studies.
We study also the non-collinear magnetic order as a possible origin of topological phases as well as the impact of the substrate on electronic band structures of monoatomic chains.
Additionally, we investigate the influence of the substrate on topological properties of magnetic chains.
In this case, we show that regardless of modeled metal atom set, the system supports a topological phase.
Therefore, the substrate plays a crucial role in the emergence of topological properties of the studied systems.

This paper is organized as follows.
First, we describe in detail the methods of investigation (Sec.~\ref{sec.methods}).
Next, we present and discuss our numerical results (Sec.~\ref{sec.num}).
Results for the isolated chains are presented in Sec.~\ref{sec.num_freestand}, 
while for the chains deposited on the substrate in Sec.~\ref{sec.num_sub}.
The latter is supplemented by the magnetic order reported in 
Sec.~\ref{sec.mo}. Finally, we summarize the results in Sec.~\ref{sec.sum}.

\section{Methods}
\label{sec.methods}

The ground state of electronic structure can be described by density
functional theory (DFT)~\cite{lejaeghere.bihlmayer.16}. Typically,
the electronic band structure is in a good agreement with experimental
data given by {\it e.g.} angle-resolved photoemission spectroscopy (ARPES).
In our study we adopted the following method of investigation:
(i) DFT calculations of electronic properties and
(ii) construction of a realistic TBM.

This allows for the comparison of the parameters obtained for bulk
crystals with the results from calculations for isolated nanowires,
{\it i.e.}, an atomic chain in the absence of substrate.
The parameters calculated for the
atomic chains are used to obtain the topological invariants for the isolated nanochains.
Next, we find the band structure for the chains deposited on a superconducting substrate, which
corresponds to a realistic situation where the orbitals of atoms from the
chain hybridize with the substrate orbitals (cf. Fig.~\ref{fig.schem}).
Finally, we derive the TBM and obtain the topological invariants
for the nanochains deposited on the Pb surface.

\subsection{Ab initio calculations}

The DFT calculations were performed using the {\sc Quantum Espresso}
code~\cite{giannozzi.baroni.09,giannozzi.andreussi.17}. The
exchange-correlation functional was calculated within the generalized
gradient approximation~\cite{perdew.chevary.92} developed by Perdew,
Burke, and Enzerhof~\cite{perdew.burke.96}. The wave functions in
the core region were evaluated using the full potential projector
augmented-wave method~\cite{blochl.94,kresse.joubert.99}.
We performed calculations in the absence and in the presence of the
spin-orbit coupling (SOC), using pseudopotentials developed in frame of
{\sc PSlibrary}~\cite{dalcorso.14}. Within the DFT calculations, we
executed a full optimization of the structural parameters for
conventional cells (for bcc and hcp structures) and primitive cells
(for an isolated chain with the vacuum layer of 10~\AA).

Additionally, to study the impact of the additional neighbors in the
chain states, we modeled a system with the substrate in approximated
form, where the chain is coupled to one layer of superconducting
substrate (containing three atoms of Pb) with vacuum layer of 10~\AA.
In calculations, we used the Monkhorst-Pack scheme~\cite{monkhorst.pack.76}
with 12$\times$12$\times$12 (12$\times$12$\times$4) {\bf k}-grid in the
case of Fe-bcc (Co-hcp) and 4$\times$4$\times$12 for 
isolated nanowires and nanowires deposited on the Pb substrate.
We have also used the cutoff for charge density with the value suggested
by using pseudopotentials increased by 100~Ry and cutoff for wave functions
with the value equals to a quarter of the charge density cutoff.

\subsection{Tight binding model}

Using the band structure obtained from the DFT calculations, we can find
the realistic TBM of the monoatomic chains in the basis of the maximally
localized Wannier functions
(MLWF)~\cite{marzari.vanderbilt.97,souza.marzari.01,marzari.mostofi.12}.
We perform this part of calculations using the {\sc Wannier90} software
\cite{mostofi.yates.08,mostofi.yates.14,pizzi.vitale.20}.
This allows for description of our system by using TBM in the form:
\begin{eqnarray}
\label{eq.ham0_re}
\mathcal{H}_{0} = \sum_{{\bm R}{\bm R'},\mu\nu,\sigma\sigma'}
T_{\mu\nu}^{\sigma\sigma'} ( {\bm R} , {\bm R}' )
c_{{\bm R}\mu\sigma}^{\dagger} c_{{\bm R}'\nu\sigma'} ,
\end{eqnarray}
where $c_{{\bm R}\mu\sigma}^{\dagger}$ ($c_{{\bm R}\mu\sigma}$) is the
creation (annihilation) operator in the MLWF basis. Here
$T_{\mu\nu}^{\sigma\sigma'} ({\bm R},{\bm R}')$ is the matrix describing
the electron hopping from orbital $\nu$ located at ${\bm R}'$ with spin
$\sigma'$ to orbital $\mu$ located at ${\bm R}$ with spin $\sigma$.
In this description, the hopping without and with spin-flip component
corresponds to the kinetic and SOC term, respectively.

When the chain is coupled to the superconductor, the superconducting 
gap $\Delta$ can be induced by the proximity effect of superconducting 
substrate (Pb in our case).
Then, our system can be described by the Hamiltonian:
\begin{eqnarray}
\mathcal{H} = \mathcal{H}_{0} + \mathcal{H}_{\rm SC},
\end{eqnarray}
where the first term denotes the ``free'' electrons (band structure),
{\it i.e.}, the Hamiltonian~(\ref{eq.ham0_re}) in momentum space:
\begin{eqnarray}
\mathcal{H}_{0} = \sum_{\bm k} H_{\mu\nu}^{\sigma\sigma'}
({\bm k)} c_{{\bm k}\mu\sigma}^{\dagger} c_{{\bm k}'\nu\sigma'} ,
\end{eqnarray}
where $H_{\mu\nu}^{\sigma\sigma'} ({\bm k}) =
\sum_{{\bm R},{\bm R}'} \exp \left[ i {\bm k} \cdot ( {\bm R}-{\bm R}' )\right]
T_{\mu\nu}^{\sigma\sigma'} ( {\bm R} , {\bm R}' )$.
The second term, describes superconductivity and can be written in the
BCS-like form:
\begin{eqnarray}
\mathcal{H}_{\rm SC} = \Delta \sum_{{\bm k}\nu}
\left( c_{-{\bm k}\nu\downarrow} c_{{\bm k}\nu\uparrow} + h.c. \right) ,
\end{eqnarray}
where $\Delta$ is half of the superconducting gap (for lead $2\Delta \sim 2.7$~meV
\cite{kittel.76,townsend.sutton.62,ji.zhang.08,ruby.heinrich.15}).
Now, $c_{{\bm k}\mu\sigma}^{\dagger}$ ($c_{{\bm k}\mu\sigma}$) is the
creation (annihilation) operator of the electron with spin $\sigma$ and
momentum ${\bm k}$ in orbital $\mu$.

\subsection{Non--trivial topological phase}

In the case of the one-dimensional hybrid semiconductor--superconductor nanowires
\cite{lutchyn.bakkers.18,mourik.zuo.12,deng.yu.12,das.ronen.12,
finck.vanharlingen.13,lee.jiang.13,deng.vaitiekenas.16,albrecht.higginbotham.16,
nichele.drachmann.17},
the phase transition from a trivial to topological phase can occur, 
when splitting of the bands given by the SOC is larger than the 
superconducting gap~\cite{sato.takahashi.09,sato.fujimoto.09,sato.takahashi.10},
\begin{eqnarray}
\mu_{B} H_{z} = \sqrt{ \tilde{\mu}^{2} + \Delta^{2} },
\end{eqnarray}
where $\mu_{B}$ is the Bohr magneton, $H_{z}$ is the magnetic field 
parallel to the nanowire, $\Delta$ is the superconducting gap, while 
$\tilde{\mu}$ is the Fermi energy computed at the bottom of the band.
In our case, the magnetic moment plays the role of the effective ``magnetic field''.
Here, it should be noted, that in contrast to the hybrid nanostructure, realization of the topological phase is given only by the intrinsic properties of the monoatomic chain, {\it e.g.} magnetic order or the position of Fermi level (which strongly depends on the type of atoms and the lattice parameters).
Therefore, it is crucial to obtain the correct TBM of studied system and our proposed solution is to use the method described in the previous section.

The topological phase can be described by a topological invariant, {\it e.g.} the winding number $w$~\cite{chiu.teo.16}. However, in our case, we describe the topological phase by the Pfaffian of the transformed Hamiltonian, which is a $\mathbb{Z}_{2}$ invariant~\cite{kitaev.01}.
This type of invariant can be defined for any system described by the Bogoliubov--de~Gennes equations~\cite{tewari.sau.12}, which is
equivalent to the Hamiltonian $\mathcal{H}$.
Because our system has the particle--hole symmetry, {\it i.e.}, ${\bm k} = 0,\pi$ are the particle--hole symmetric points in the Brillouin zone~\cite{more.balents.07}, the Pfaffian is given by~\cite{tewari.sau.12}:
\begin{eqnarray}
\label{eq.topo_inv}
\mathcal{Q} = \text{sgn} \left[ \frac{ \text{Det}(A({\bm k}=\pi)) }
{ \text{Det}(A({\bm k}=0)) } \right] = ( -1 )^{w} .
\end{eqnarray}
Here, $A({\bm k})$ denotes the element of Hamiltonian matrix in the block off-diagonal form~\cite{schnyder.ryu.08}, which can be derived from the unitary transformation $\mathcal{U}$~\cite{ryu.schnyder.10}:
\begin{eqnarray}
\mathcal{U} \mathcal{H} \mathcal{U}^{\dagger} = \left(
\begin{array}{cc}
0 & A ({\bm k}) \\
A^{T} (-{\bm k}) & 0
\end{array}
\right) ,
\end{eqnarray}
where $A_{\sigma\sigma'}^{\mu\nu} ({\bm k}) = H_{\sigma\sigma'}^{\mu\nu} ({\bm k}) + \Delta \delta_{\bar{\sigma}\sigma'} \delta_{\mu\nu}$.
The topological phase is realized when $\mathcal{Q} = -1$.

\begin{figure}[!t]
\includegraphics[width=0.95\linewidth]{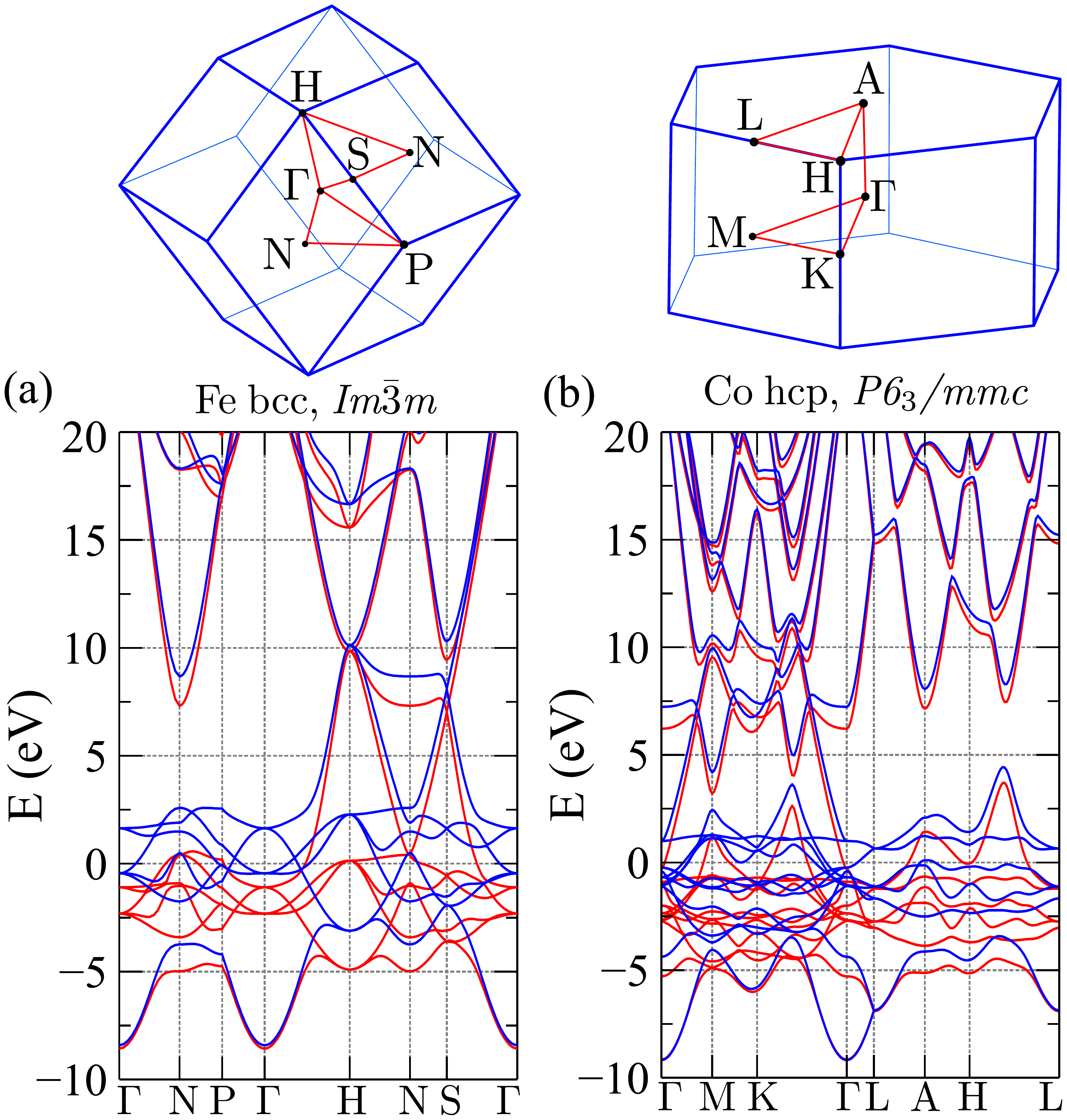}
\caption{
First Brillouin zone and band structures of the Fe bcc and Co hcp
crystals. Results in the absence of the SOC.
Red and blue colors denote the states with spin $\uparrow$ and
$\downarrow$, respectively. Fermi level is located at zero energy.
\label{fig.band_bulk}
}
\end{figure}

\section{Numerical results}
\label{sec.num}

We start from a short description of the Fe bcc and Co hcp bulk systems.
From the DFT self--consistent calculations, we find that the Fe bcc (Co hcp) structure have magnetic moments equal to 2.1988~$\mu_{B}$ (1.6693~$\mu_{B}$) and lattice constant of 2.4512~\AA\ (2.4881~\AA). For the optimized systems, we find the electronic band structures (Fig.~\ref{fig.band_bulk}).
In both cases, the $3d$ orbitals are accumulated around the Fermi level, while the rest of states (unoccupied $4p$ states) are located far above the Fermi level (approximately above 7.5~eV).

\subsection{Isolated chains}
\label{sec.num_freestand}

\begin{figure}[!t]
\includegraphics[width=0.95\linewidth]{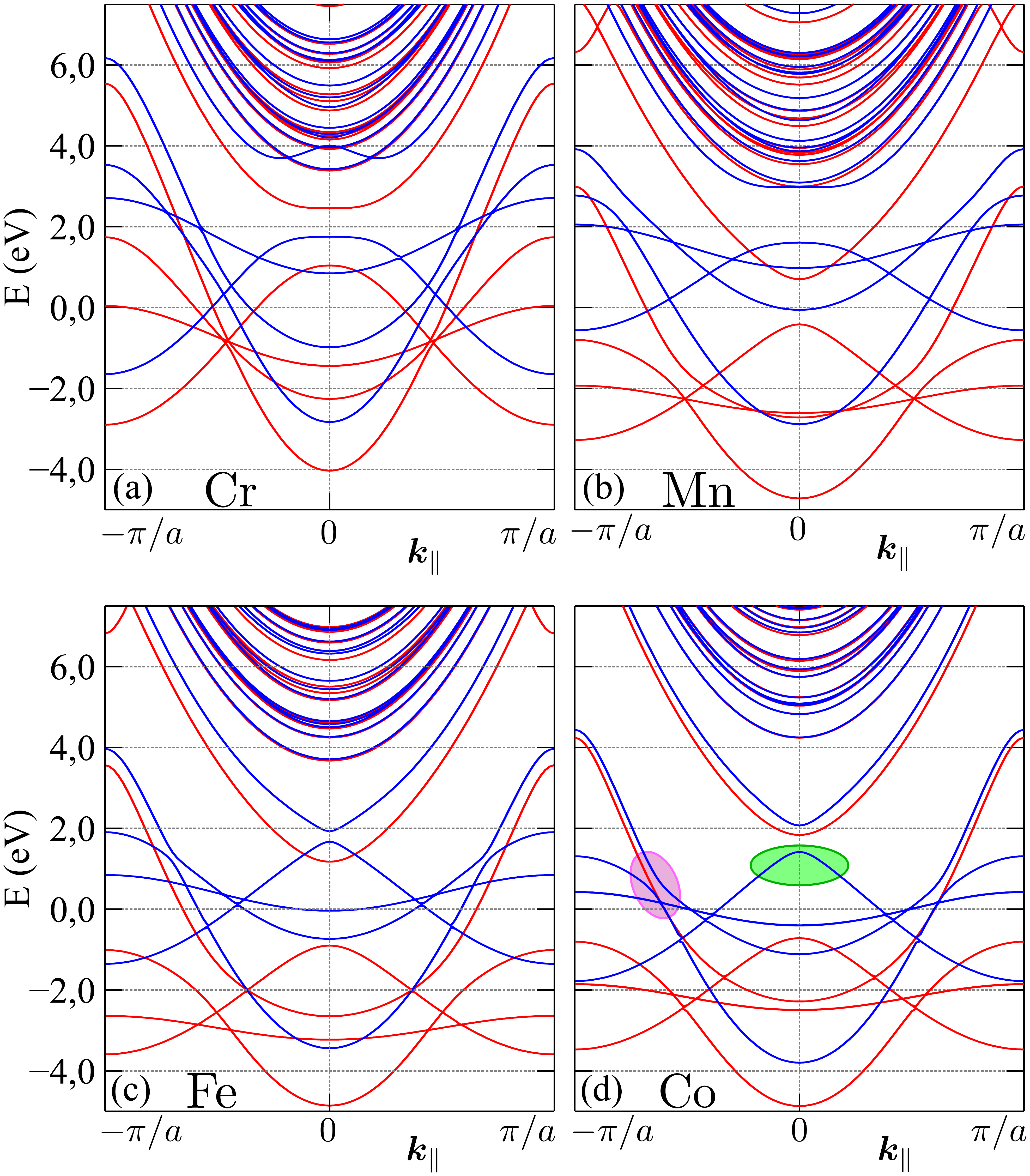}
\caption{
Electron band structures of Cr, Mn, Fe, and Co nanowires as labeled).
Results obtained at absence of the SOC.
Red and blue colors denote states with the spin $\uparrow$ and $\downarrow$, 
respectively. Fermi level is located at zero energy.
\label{fig.band_nw}
}
\end{figure}

Now we discuss the results for the isolated magnetic chains.
Here, we performed the volume relaxation of one magnetic atom with the
15~\AA \  of vacuum in $\hat{x}$ and $\hat{y}$ directions (chain is
aligned along the $\hat{z}$ direction). From this, we find the distances
between atoms in the isolated
nanowires (see Tab.~\ref{tab.stale_sieciowe}).
The obtained distances in both Fe and Co chains are approximately~0.2~\AA
\ smaller than those in the bulk materials, while magnetic moments are
larger.
Modification of these two quantities must have a substantial impact on
the parameters of the model describing the atomic chains. It is clearly
visible in the band structures of the isolated
chains (see Fig.~\ref{fig.band_nw}).
We observe a strong shift of the $p$-orbital states to lower energies
(cf. Fig.~\ref{fig.band_bulk} and Fig.~\ref{fig.band_nw}, states
initially located above 10~eV are shifted to energies around 4~eV).
This leads to the strong hybridization between these states with the
$3d$ levels. More importantly, one additional band crosses the Fermi
level. In consequence, isolated monoatomic chains cannot be described
not only by a simple single-orbital tight binding model, but even by
a model incorporating as much as ten $3d$ orbitals.

\begin{table}[!b]
\caption{
Distances between atoms (in~\AA) and magnetic moments (in $\mu_{B}$)
in the 
isolated chains.
\label{tab.stale_sieciowe}
}
\begin{ruledtabular}
\begin{tabular}{ l c c c c }
$3d$ element & Cr & Mn & Fe & Co \\
\colrule
distance & 2.07 & 2.30 & 2.23 & 2.15 \\ 
mag. mom. w/o SOC & 1.94 & 3.50 & 2.89 & 2.05 \\
mag. mom. w/ SOC & 1.78 & 3.55 & 2.94 & 2.05
\end{tabular}
\end{ruledtabular}
\end{table}

The shapes of the obtained bands associated with $3d$ orbitals are
approximately given as a cosine-like function of momentum (see Fig.
\ref{fig.band_nw}), which is typical for a one--dimensional chain.
However, when $3d$ states hybridize with other orbitals, a relatively
large ``deformation'' of this shape (marked by green circle) takes
place. Therefore, the band structure cannot be approximated by the
dispersion relation of a simple one--dimensional lattice anymore.
The other consequence is an avoided crossings behavior of the
hybridized bands (marked by pink circle).

\begin{figure}[!b]
\includegraphics[width=0.95\linewidth]{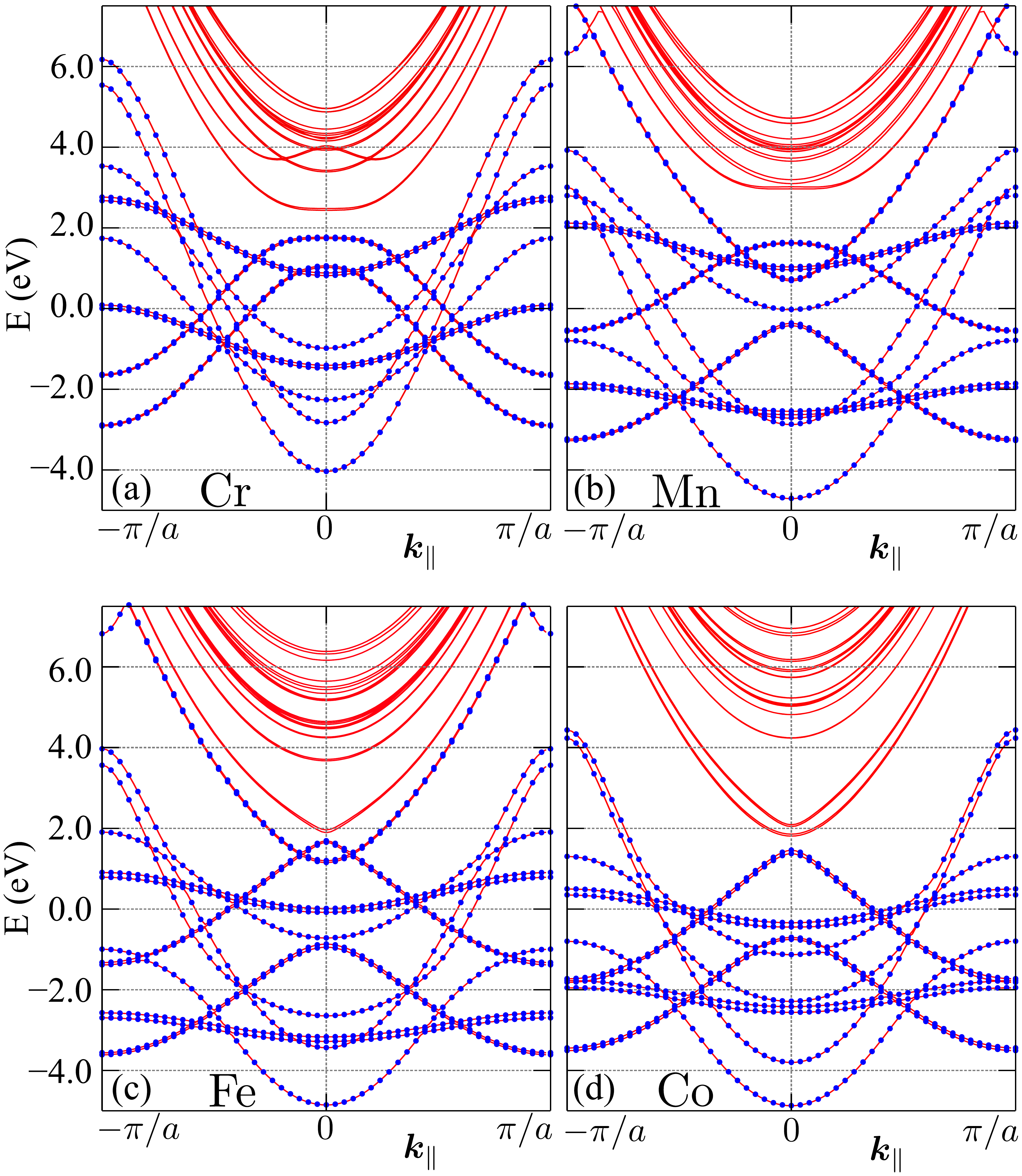}
\caption{
The same as in Fig.~\ref{fig.band_nw} in the presence of the SOC and 
magnetic moment parallel to the nanowire.
Solid red lines and blue dots correspond to band structures obtained 
from the DFT and TBM calculations, respectively.
Fermi level is located at zero energy.
\label{fig.band_nw_soc}
}
\end{figure}

Introduction of the spin-orbit coupling in the calculations does not change the results qualitatively.
As usually, the band degeneracy is lifted thanks to spin--orbit coupling, however, the shape of band  dispersion is not influenced. The magnetic moments found in non-collinear calculations have approximately similar values, independently of its direction. Still, the largest splitting of the bands can be found when the magnetic moments are parallel to the nanowire.
Even though splitting of the bands due to the spin-orbit coupling depends on the atomic mass~\cite{shanavas.popovic.14}, it is much smaller for isolated nanowires than in the bulk.
Here it should be mentioned, that in the previous studies of isolated Fe and Co free--standing chains~\cite{nadjperge.drozdov.14,ruby.heinrich.17}, a value of the spin-orbit coupling was overestimated.

In conclusion, a difference in the distance between atoms in the chain and in the bulk, as well as a reduced number of neighboring atoms, leads to severe modification of most of system parameters, {\it e.g.} hoping integrals, magnetic moments, or spin--orbit coupling.
Additionally, $p$--type orbitals  play an important role in a proper description of the isolated chains.

\paragraph*{Realization of a non--trivial topological phase.---} 
In the previous studies of the magnetic monoatomic chains,  in order to describe isolated nanowires capable of hosting the topologically non--trivial phase, the Slater-Koster tight-binding model parameters for bulk were used~\cite{nadjperge.drozdov.14,ruby.heinrich.17}.
In the case of the Fe chain, the hopping integral values were taken for the nearest-neighbor distance of the bulk Fe (bcc, {\it Im$\bar{3}$m}, Space group: 229), which is 2.383~\AA~\footnote{see {\it Supplementary Materials} for Ref.~\cite{nadjperge.drozdov.14}.}.
Similarly, in the case of the Co chain, the hopping integral was
calculated for the nearest neighbor distance of the bulk Co
(hcp, {\it P6$_{3}$/mmc}, Space group: 194) with $a=2.486$~\AA~\footnote{see {\it Supporting Information} for Ref.~\cite{ruby.heinrich.17}.}.
Taking into account strong modifications of the band structure in the isolated chains (see Figs.~\ref{fig.band_nw} and \ref{fig.band_nw_soc}), in particular a different number of bands crossing the Fermi level, such assumptions can lead to incorrect conclusions regarding the existence of the non--trivial topological phase.
Moreover, in general case, fitting the band structure with the Slater-Koster parameters can give incorrect results in comparison with the {\it ab initio} band structure~\cite{roldan.chirolli.17}.

To precisely describe the band structures of isolated chains, we found
the TBM in the MLWF based on the DFT calculations
(cf. Fig.~\ref{fig.band_nw_soc}). In a general case, the TBM model
describing our system around the Fermi level is mostly composed of
$d$-like orbitals (typical for transition metals). However, contrary to
the bulk models, additional $p$-like orbitals should be included in the
model.

Next, by employing these models, we calculate the topological number $\mathcal{Q}$ given by Eq.~(\ref{eq.topo_inv}).
Additionally, we assume that the superconducting gap induced by the proximity effect in nanowires is equal 3~meV (which is close to the experimental result of 2.7~meV).
That small, qualitative, difference does not change results, which are presented in Tab.~\ref{tab.topo}.
As we mentioned before, the topological phase can be realized only when $\mathcal{Q}=-1$.
From our calculations we can conclude that the topological phase can be induced only in the Mn and Co nanowires.

\begin{table}[!b]
\caption{
\label{tab.topo}
Signs of the Pfaffians in the time-reversal invariant momenta and values of the topological number $\mathcal{Q}$.
Results obtained for the isolated chains.
}
\begin{ruledtabular}
\begin{tabular}{ccccc}
          & \multicolumn{4}{c}{$3d$ element} \\
${\bm k}$ & Cr & Mn & Fe & Co \\ \hline \hline
$0$ & $+$ & $+$ & $-$ & $+$ \\
$\pi$ & $+$ & $-$ & $-$ & $-$ \\
\hline
$\mathcal{Q}$ & $+1$ & $-1$ & $+1$ & $-1$
\end{tabular}
\end{ruledtabular}
\end{table}

\begin{table*}[t]
\caption{
\label{tab.ncl}
Comparison of the energies between non-collinear magnetic orders with magnetic moments lying in the plane containing the chain ($E_{\parallel}$) and in the plane perpendicular to the chain ($E_{\perp}$), $\delta E = E_{\parallel} - E_{\perp}$, which plays the role of the magnetic anisotropy energy.
The angle of the magnetic moment rotation in space is given by $\varphi$, which depends on number of the atoms in magnetic unit cell $n$, {\it i.e.}, $\varphi = 2\pi/n$.
Results obtained in the presence of SOC, in eV per atom.
}
\begin{ruledtabular}
\begin{tabular}{ccccccccc}
$\varphi$ & \multicolumn{2}{c}{Cr} & \multicolumn{2}{c}{Mn} & \multicolumn{2}{c}{Fe} & \multicolumn{2}{c}{Co} \\
    & $E_{\parallel}-E_{0}$  & $E_{\perp}-E_{0}$     & $E_{\parallel}-E_{0}$  & $E_{\perp}-E_{0}$    & $E_{\parallel}-E_{0}$  & $E_{\perp}-E_{0}$     & $E_{\parallel}-E_{0}$  & $E_{\perp}-E_{0}$  \\ \hline \hline
$0$ & $-6.5\times10^{-4}$ & $4.2\times10^{-6}$ & $-1.538$ & $-1.538$ & $-1.101$ & \fbox{$-1.108$} & \fbox{$-0.491$} & $-0.470$ \\
   & \multicolumn{2}{c}{$\delta E = -6.5\times 10^{-4}$} & \multicolumn{2}{c}{$\delta E = 7.8\times 10^{-7}$} & \multicolumn{2}{c}{$\delta E = 3.4\times 10^{-6}$} & \multicolumn{2}{c}{$\delta E = -0.020$} \\
$\pi$ &  \fbox{$-1.057$} & $-1.056$ & $-1.794$ & $-1.794$ & $-0.800$ & $-0.799$ & $-0.074$ & $-0.025$ \\
   & \multicolumn{2}{c}{$\delta E = -8.3\times 10^{-4}$} & \multicolumn{2}{c}{$\delta E = 3.0\times 10^{-5}$} & \multicolumn{2}{c}{$\delta E = -1.3\times 10^{-3}$} & \multicolumn{2}{c}{$\delta E = -0.049$} \\
$2\pi/3$ & $-0.582$ & $-0.582$ & $-1.828$ & \fbox{$-1.828$} & $-0.923$ & $-0.923$ & -0.144 & $-0.149$ \\
   & \multicolumn{2}{c}{$\delta E = -4.3\times 10^{-8}$} & \multicolumn{2}{c}{$\delta E = 1.3\times10^{-4}$} & \multicolumn{2}{c}{$\delta E = -1.7\times 10^{-6}$} & \multicolumn{2}{c}{$\delta E = 0.006$} \\
$\pi/2$  & $-0.420$ & $-0.420$ & $-1.766$ & $-1.766$ & $-1.047$ & $-1.015$ & $-0.310$ & $-0.318$ \\
   & \multicolumn{2}{c}{$\delta E = 0.0$} & \multicolumn{2}{c}{$\delta E = 3.2\times 10^{-8}$} & \multicolumn{2}{c}{$\delta E = -0.032$} & \multicolumn{2}{c}{$\delta E = 0.008$} \\
   $2\pi/5$  & $-0.914$ & $-0.366$ & $-1.707$ & $-1.707$ & $-1.089$ & $-1.089$ & $-0.460$ & $-0.349$ \\
   & \multicolumn{2}{c}{$\delta E = -1.6 \times 10^{-7}$} & \multicolumn{2}{c}{$\delta E = -4.0 \times 10^{-7}$} & \multicolumn{2}{c}{$\delta E = -2.8 \times 10^{-7}$} & \multicolumn{2}{c}{$\delta E = 0.041$}
\end{tabular}
\end{ruledtabular}
\end{table*}

Information about the realization of a non--trivial phase, can be attained from the number of the bands crossing the Fermi level within a half of the Brillouin zone~\cite{hasan.kane.10}.
If number of crossing bands is odd then the realization of the Majorana bound states at the ends of the finite nanowire is expected.
In the case of the Fe chain described by model with bulk parameters~\cite{nadjperge.drozdov.14}, this number was almost always odd, making the presence of Majorana bound states at the ends of the chains almost guaranteed.
On the other hand, for the Co chain with the SOC number of crossings was even~\cite{ruby.heinrich.17}.
From our results (cf.~Fig.~\ref{fig.band_nw_soc}), even number of band crossing the Fermi level within half of the Brillouin zone are realized in the Cr and Fe nanowire. These analyzes yield comparable results to those obtained from $\mathcal{Q}$, supporting the hypothesis about realization of the non--trivial topological phase in Mn and Co nanowires.

The obtained results contradict the previous studies of Fe and Co chains~\cite{nadjperge.drozdov.14,ruby.heinrich.17}, where topological phase was found only in the Fe chain.
Here, we must keep in mind, that such results are extremely sensitive to parameters used in calculations.
For instance, distances between atoms can modify the band structure by changing the overlap between orbitals.
Similarly, the number of occupied bands for momentum $0$ or $\pi$ can strongly depend on a position of the Fermi level -- this should be important in the case of Mn, Fe or Cr isolated chains, where bands at momentum $0$ or $\pi$ are located close to the Fermi level.
In such case, a small modification of the hopping integrals, as well as the SOC or magnetic moments, can change a value of $\mathcal{Q}$ drastically and in consequence, topological properties of the system.

\paragraph*{Non-collinear magnetic moments.---}
A topologically non--trivial phase is not exclusive to a ferromagnetic chain --- in some situations topological effects can also be induced by non-collinear magnetic moments.
In the chain of one-orbital magnetic ``atoms'', the spiral order can minimize the free energy of the system leading to the emergence of a topological phase~\cite{vazifeh.franz.13,reis.marchand.14}.
In this situation, the Majorana quasiparticles can be found at the ends of the chain~\cite{braunecker.simon.13,pientka.glazman.13,klinovaja.stano.13,kim.cheng.14}.
This is possible due to the fact that the spiral magnetic order leads to the same effects as the SOC together with the external magnetic field~\cite{braunecker.japaridze.10,klinovaja.loss.12}.

In a more realistic situation of a multi-orbital chain, the description using only a simple model may not be sufficient~\cite{li.chen.14}.
However, the DFT calculations allow for a comparison of the energies of the chains with different non-collinear magnetic orders
(Tab.~\ref{tab.ncl}).
From this comparison, we can find the order which minimizes the energy of the system (values in the box).
As we can see, in the case of Fe and Co atoms, the ferromagnetic order is more favorable.
In Fe chain, the magnetic moment should be perpendicular to the chain, in contrast to the parallel moment in Co chain.
Interestingly, in Cr chain, the antiferromagnetic order is the most stable one, while in Mn chain a chiral order with the $2\pi/3$ period is the lowest energy state.
Thus, we do not expect the non-collinear magnetic order as a probable source of the Majorana quasiparticles in Fe and Co magnetic chains.

\begin{figure}[!b]
\centering
\includegraphics[width=\columnwidth]{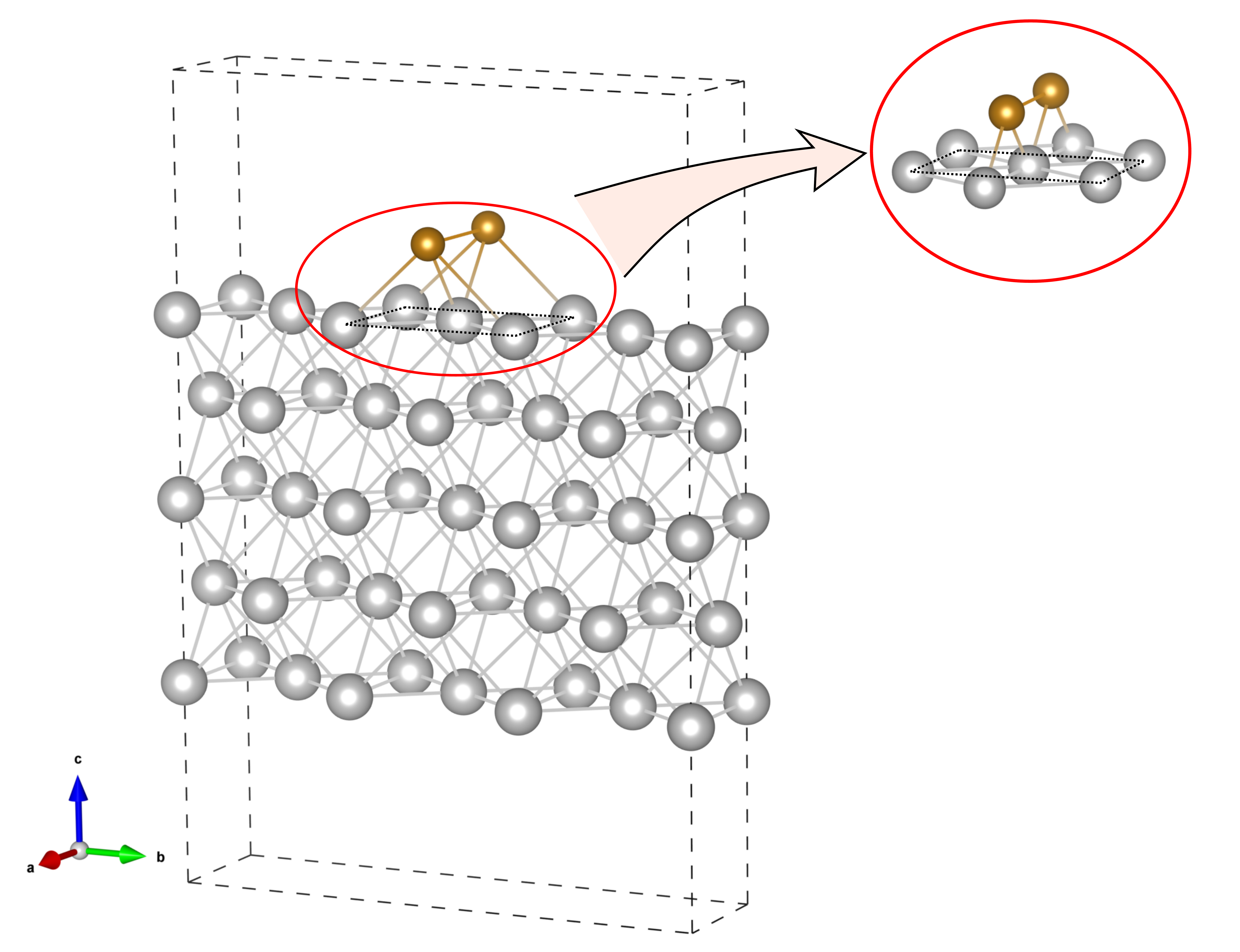}
\caption{
Schematic representation of the unit cell (black dashed line) in the case of a monoatomic chain deposited on the Pb(110) surface.
To check the influence of the neighboring atoms on the band structure of the chain, for simplicity, we consider a system in the form presented in the inset (containing one transition metal atom and three lead atoms in the unit cell).
The image was rendered using {\sc Vesta} software~\cite{vesta}.
\label{fig.sub_uc}
}
\end{figure}

\subsection{Chains deposited on the substrate}
\label{sec.num_sub}

Now, we will discuss the results obtained for the chains deposited on a substrate---for the system presented in Fig.~\ref{fig.schem}.
In order to simulate the measurements described in Refs.~\cite{nadjperge.drozdov.14} and~\cite{ruby.heinrich.17}, we take the Pb(110) surface as the substrate.
To simplify the band structure and make it more readable, in the calculations we have used a system shown in the inset in Fig.~\ref{fig.sub_uc}: containing one transition metal atom and three lead atoms in the unit cell.
In the first approximation such a system can help us to describe the influence of neighboring Pb atoms on the transition metal chain.

\begin{figure}[!pt]
\centering
\includegraphics[width=0.94\linewidth]{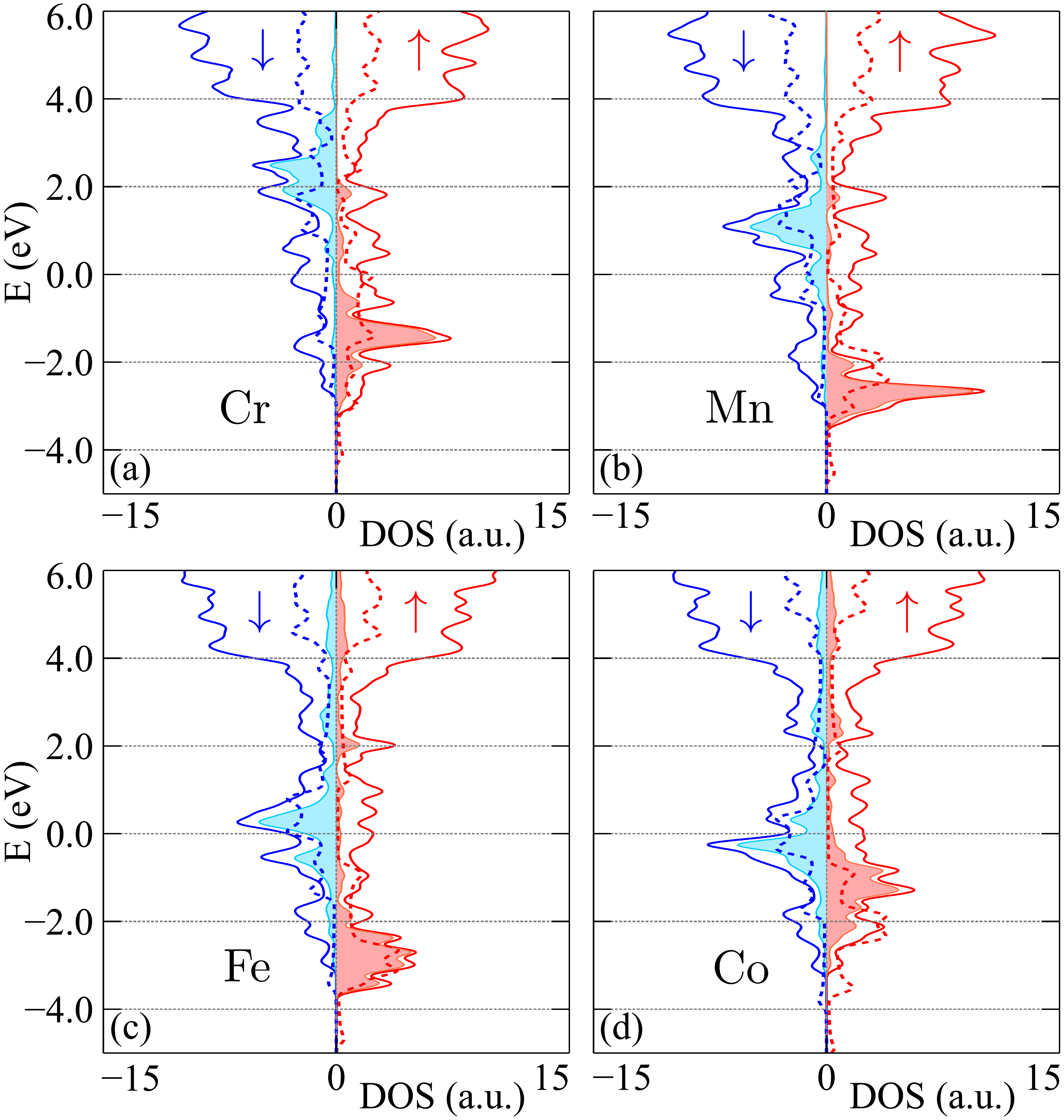}
\caption{
Comparison of the electronic density of states (DOS) for chains (as labeled) isolated and deposited on the substrate in the approximated case (as shown in Fig.~\ref{fig.sub_uc}).
Red and blue lines denote states with $\uparrow$ and $\downarrow$ spin, respectively.
Solid areas show contribution of the deposited atoms.
Fermi level is located at zero energy.
\label{fig.dos}
}
\end{figure}

\begin{figure}[!pb]
\centering
\includegraphics[width=0.94\linewidth]{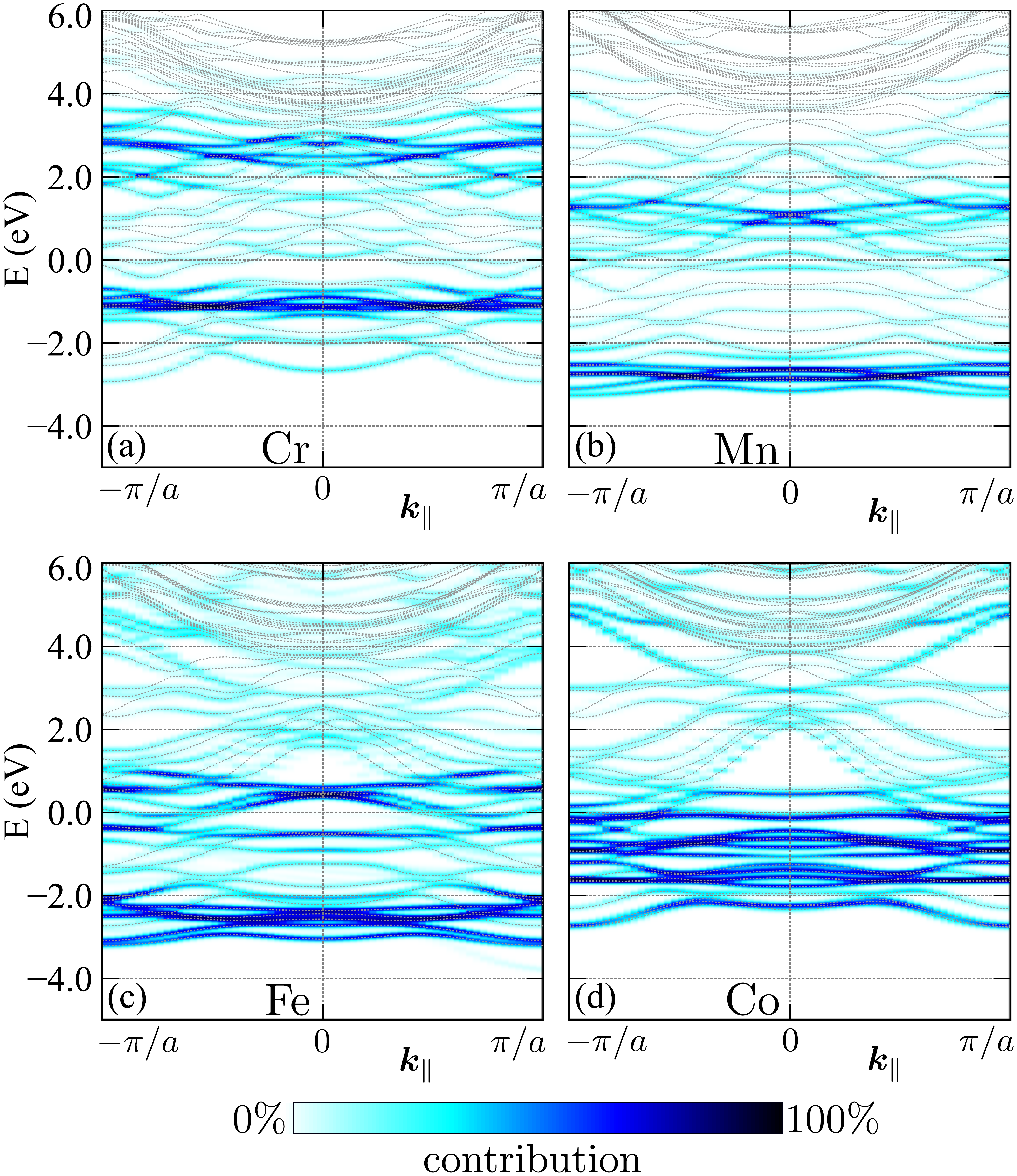}
\caption{
Electronic band structures of nanowires (as labeled) deposited on the substrate (as shown in Fig.~\ref{fig.sub_uc}).
Results in the presence of the SOC.
Color in background shows contribution of the deposited atoms.
Fermi level is located at zero energy.
\label{fig.sub_band_aprox_soc}
}
\end{figure}

Figure \ref{fig.dos} presents the density of states (DOS)  in the case of the isolated chain (dashed line) and the chain deposited on the substrate (solid line).
In the latter case, contributions of the deposited atoms to the total DOS are shown by solid-colored areas.
Comparing to the isolated chains, all $3d$-orbital bands are  modified and become narrower.
The weakest effect of the substrate is observed for the Fe chain, where the positions of spin-up states very well correspond to those in the isolated chain.
Analyzing the electron DOS, we can also explain the modification of the magnetic moments induced by the substrate.
In the case of Cr and Mn, the atoms are nearly fully spin polarized (all $\downarrow$ states are above the Fermi level, while $\uparrow$ below).
In contrast, a small magnetic moment of Co results from the shift of the $\downarrow$ states to energies below the Fermi level.

Deposition of the chains on the substrate changes the system symmetry and allows for the hybridization between the orbitals in the chain and the substrate.
These properties lead to the modification of the band structure of the studied system (Fig.~\ref{fig.sub_band_aprox_soc}).
To increase the readability of the band structure, the projection of the states into transition metal atoms are shown by colors in the background.
From the comparison of the band structures of the isolated chain and the deposited chain, see Figs.~\ref{fig.band_nw} and
\ref{fig.sub_band_aprox_soc}, respectively), we can find the influence of the substrate on the chain bands.
The bands associated with transition metal atoms have narrower bandwidth with respect to the isolated nanowire.
This is equivalent to the modification of the hopping integrals between the atomic orbitals.

\begin{figure}[!b]
\centering
\includegraphics[width=\linewidth]{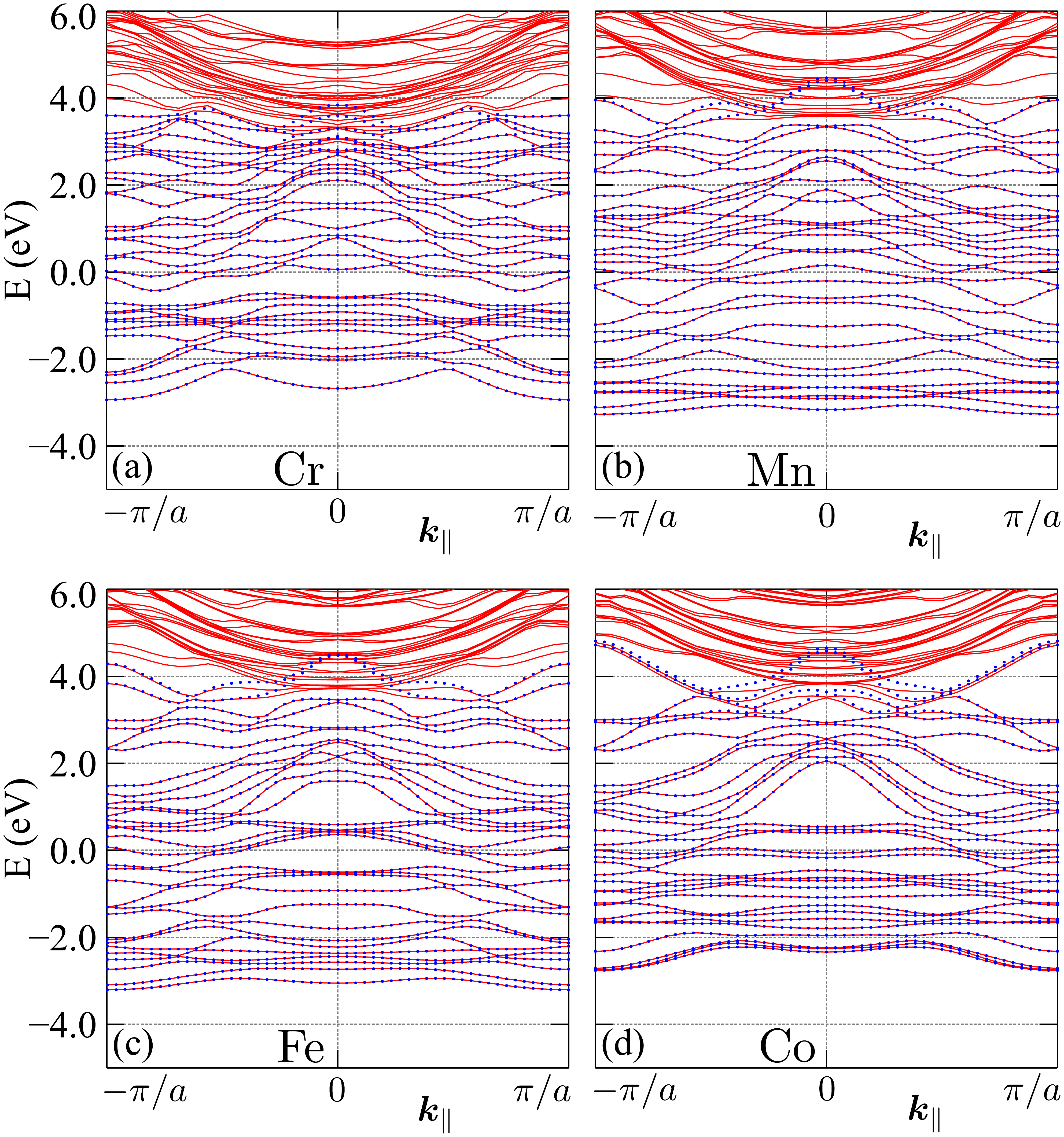}
\caption{
Comparison of the band structures obtained from the DFT and TBM calculations (red lines and blue dots, respectively).
Results for nanowires deposited on Pb(110) surface (as shown in Fig.~\ref{fig.sub_uc}).
Fermi level is located at zero energy.
\label{fig.sub_tbm}
}
\end{figure}

\paragraph*{Non--trivial topological phase.---}
Now, we perform the analysis of existence of the topological phase in the chains deposited on the Pb surface.
Similarly like in the previous case, in the first step, we found the TBM in MLWF based on the DFT calculations (cf.~Fig.~\ref{fig.sub_tbm}).
In contrast to the models of isolated chains, here, TBM are based on $3d$ orbitals of transition metals and $6p$ orbitals of Pb.
In total, our models take into account 30 orbitals and reproduce band structure around the Fermi level very well.
As we can see, in every case, around $4$~eV, we can observe strong interplay between orbitals included in the model and those excluded (at the $\Gamma$ point).
For the developed models, we calculate topological number $\mathcal{Q}$ 
(results are included in Tab.~\ref{tab.topo_sub}).
Surprisingly, in contrast to the isolated chains, topological 
phase is supported in every case of the chains deposited on the substrate.
These results show an important role of the substrate in stabilization 
of the topological phase.

\begin{table}[!t]
\caption{
\label{tab.topo_sub}
Signs of the Pfaffians in the time-reversal invariant momenta and 
values of the topological number $\mathcal{Q}$. Results for nanowires 
deposited on Pb(110) surface (as shown in Fig.~\ref{fig.sub_uc}).}
\begin{ruledtabular}
\begin{tabular}{ccccc}
          & \multicolumn{4}{c}{$3d$ element} \\
${\bm k}$ & Cr & Mn & Fe & Co \\ \hline \hline
$0$ & $-$ & $+$ & $-$ & $-$ \\
$\pi$ & $+$ & $-$ & $+$ & $+$ \\
\hline
$\mathcal{Q}$ & $-1$ & $-1$ & $-1$ & $-1$
\end{tabular}
\end{ruledtabular}
\end{table}

\begin{table}[!b]
\caption{
Magnetic moments (in $\mu_{B}$) in the monoatomic chain deposited
on the Pb surface (as shown in~Fig.~\ref{fig.sub_uc}). Results in the absence and in the presence of SOC.
\label{tab.mag_sub}
}
\begin{ruledtabular}
\begin{tabular}{ l c c c c }
$3d$ element & Cr & Mn & Fe & Co \\
\colrule
mag. mom. without SOC & 4.16 & 3.91 & 2.78 & 1.08 \\
mag. mom. \hspace{1mm} with \hspace{1mm} SOC & 4.18 & 3.99 & 2.77 & 1.12 \\
\end{tabular}
\end{ruledtabular}
\end{table}

\subsection{Magnetic order}
\label{sec.mo}

Now, we will shortly discuss the magnetic order of the chains deposited on the substrate.
Coupling the chain to the Pb atoms leads to the increase of the magnetic moments only in the case of Cr atom (cf. Tab.~\ref{tab.stale_sieciowe} and~\ref{tab.mag_sub}).
One should also notice the strong suppression of the magnetic moment in the Co chain.

As we mentioned in previous paragraphs, non-collinear magnetic moments can be a source of the topological phase.
This type of magnetic structures can be stabilized by the conduction-electron-mediated Ruderman--Kittel--Kasuya--Yosida (RKKY) interaction~\cite{zhou.wiebe.10}.
For instance, early studies of the RKKY mechanism shown that the FM order in Fe chains is unstable~\cite{anderson.suhl.59}.
However, more recent theoretical studies allow for existence of the FM, AFM or non-collinear magnetic orders~\cite{schubert.mokrousov.11,tanveer.riozdiaz.13}, depending on the system parameters.
Additionally, it was experimentally shown, that the small cluster of magnetic atoms can exhibit AFM instability~\cite{loth.baumann.12,yan.malavolti.17}.
Similarly, the non-collinear magnetic order in Fe double-chain~\cite{menzel.mokrousov.12} have been reported.
Here, the RKKY interaction can lead to stabilization of the non-collinear magnetic orders~\cite{hermenau.brinker.19}.
On the other hand, a role of the Dzyaloshinskii--Moriya interaction (DMI) can be important~\cite{steinbrecher.rausch.18} in the case of the chain deposited on a substrate, due to an interface between these two systems~\cite{belabbes.bihlmayer.16}.
Additionally, the strength of DMI can depend on the position of a chain with respect to a surface~\cite{schweflinghaus.zimmermann.16}.
In conclusion, both types of interactions, as well as a its mutual interplay, have important role in adjusting the interatomic iron distance, which enables tailoring of the rotational period of the spin-spiral~\cite{steinbrecher.rausch.18}.
The inclusion of these interactions in calculations with the substrate may change the ground state
and support the topological phase. However, unfortunately such long-range interactions like RKKY cannot be included in our calculations due to a small size of the cell.

Here, we should also remember about a general form of the interaction between electrons inside $3d$ atoms chains~\cite{oles.83}, which include intra- and inter-orbital Coulomb repulsions, as well as the Hund’s exchange and the pair hopping term.
The existence of strong magnetic moments in $3d$ transition metals cannot be correctly captured within the single-band model. From this, the stabilization of the 
topological phase should depend not only on the Hund's exchange in 
partly filled $3d$ orbital states~\cite{Ole84,Sto90}, but also on 
other interactions which have negative impact on the emergence of \mbox{MBS~\cite{ng.15,katsura.schuricht.15,wieckowski.ptok.19}}.

\section{Summary}
\label{sec.sum}

In summary,
Majorana quasiparticles constitute a very interesting concept of particles, which are indistinguishable from their antiparticles. One of the many platforms in which we expect the emergence of bound states with such properties are systems of magnetic atomic chains deposited on a surface of the conventional superconductor~\cite{nadjperge.drozdov.13,li.chen.14,peng.pientka.15,pawlak.silas.19}.
In this paper, we studied the topological phase of the $3d$ 
transition metal chains, in form of 
 (i)~isolated chains and
(ii) chains deposited on the Pb surface.

In Sec.~\ref{sec.num_freestand}, we discussed the free--standing chains which should be treated as a first step in a theoretical description of the experimental system.
Previous studies of the $3d$ transition metal chains~\cite{nadjperge.drozdov.14,ruby.heinrich.17}, were based on the Slater-Koster parameters of the bulk  systems~\cite{slaterkoster}.
Unfortunately, such an approach does not correctly describe the physical properties of the free--standing chains, mostly due to a different number of neighboring sites in chains comparing to a bulk.
Additionally, different distances between atoms in both systems lead to 
the strong modification of the band structure ({\it i.e.}, hopping integrals between orbitals).
To verify this, we developed the tight binding model in the Wannier orbitals, based on the {\it ab initio} (DFT) band structures.
We have shown that in the case of isolated chains, the additional band crossing the Fermi level exists, which cannot be captured by the simple tight binding model derived from the bulk electronic structure.
Using the obtained tight binding model, we also calculated the topological quantum number.
Hence, we concluded that in the case of isolated chains, the non--trivial topological phase can exist only in the Mn and Co chains.
These results are in opposition to the previous studies of the Fe and Co chains~\cite{nadjperge.drozdov.14,ruby.heinrich.17}, where non--trivial topological phase was reported only in the case of the iron chain.

Next, we performed similar analysis for the chains deposited on the Pb surface (Sec.~\ref{sec.num_sub}).
In this case, we studied the impact of the substrate on topological properties of the system.
The interplay between the atoms of the substrate and the chains leads to strong modifications of the electronic properties of the chains.
It is clearly visible in the band structure projected onto the chain atoms, as well as in the density of states.
Here, we also developed a tight binding model of this system to calculate the topological index.
In contrast to the isolated chains where the influence of the substrate is described only by one parameter (superconducting gap), the incorporation of the surface states in the system leads to the emergence of the non-trivial phase, regardless of the used transition metal.
Finally, we have shown in Sec. \ref{sec.mo} that the magnetic order 
in the chains deposited on the substrate is a subtle problem and the 
final order depends on the electronic filling which decides about the 
RKKY interaction.

We hope that our studies will provide a substantial information about 
the properties of magnetic chains and will stimulate further research
in this field. Our findings demonstrate a significant influence of the 
substrate on topological properties of the magnetic chains. Therefore, 
we conclude that a substrate constitutes a crucial part for a correct 
description of nanosystems and should be included in future studies 
based on {\it ab initio} methods.

\begin{acknowledgments}

We thank Pascal Simon and Wojciech Tabi\'{s} for inspiring discussion 
and valuable comments.
A. P. is grateful to Laboratoire de Physique des Solides 
(CNRS, Universit\'{e} Paris-Sud) for hospitality during a part of the 
work on this project.
This work was supported by the National Science Centre (NCN, Poland) 
under Projects:
2018/31/N/ST3/01746 (A. K.), 
2017/25/B/ST3/02586 (P. P.), 
2016/23/B/ST3/00839 (A. M. O.), 
and
2016/23/B/ST3/00647 (A. P.). 
A. M. O. is grateful for the Alexander von Humboldt Foundation Fellowship (Humboldt-Forschungspreis).

\end{acknowledgments}

\bibliography{biblio}

\end{document}